\documentclass[prd,showpacs,nofootinbib,preprint,11 pt]{revtex4}
\usepackage{graphicx,color}
\begin{document}

\title{Weak productions of new charmonium  in semi-leptonic decays of $B_c$ }

\author{Yu-Ming Wang}
\author{Cai-Dian L\"{u} }


\affiliation{Institute of High Energy Physics, P.O. Box
 918(4), Beijing 100049, China}

\vspace*{1.0cm}

\date{\today}
\begin{abstract}
We study the weak productions of novel heavy mesons, such as
$\eta_c^{\prime}$, $h_c$, $h_c^{\prime}$, $\chi_{c0}^{\prime}$,
$X(3940)$, $Y(3940)$, $X(3872)$, and $Y(4260)$,  in the
semi-leptonic $B_c$ decays.  Since there is still no definite answer
for the components of $X(3940)$, $Y(3940)$, $X(3872)$, $Y(4260)$ so
far, we will assign them as excited charmonium states with the
possible quantum numbers constrained by the current experiments. As
for the weak transition form factors, we calculate them in the
framework of light-cone QCD sum rules approach, which is
 proved to be a powerful tool to deal with the non-perturbative
hadronic matrix element. Our results indicate that different
interpretations of $X(3940)$ can result in remarkable discrepancy of
the production rate in the $B_c$ decays, which would help to clarify
the inner structure of the $X(3940)$ with the forthcoming LHC-b
experiments. Besides, the predicted large weak production rates of
$X(3872)$ and $Y(3940)$ in $B_c$ decays and the small semi-leptonic
decay rate for $B_c \to Y(4260)$ all depend on their quantum number
$J^{PC}$ assignments. Moreover, the $S-D$ mixing of various vector
charmonium states in the weak decay of $B_c$ is also discussed in
this work. The future experimental measurements of these decays will
test the inner structures of these particles, according to our
predictions here.

\end{abstract}

 \maketitle

\newpage

\section{Introduction}

A number of new  hidden charm states were observed recently by
experiments, such as $X(3872)$, $X(3940)$, $Y(3940)$, $Z(3930)$ and
$Y(4260)$ \cite{Belle,X(3940),Y(3940),Z(3930),Y(4260) BaBar}.  Their
quark structures are still not fully understood \cite{R. Faccini}.
  In particular, the $X(3872)$, which exhibits various impenetrable
aspects, is labeled as the poster boy of the new heavy hadrons
\cite{swanson}.
 Although the quantum numbers $J^{PC}=1^{++}$ of
$X(3872)$ are strongly favored by the experiments, there is not a
definite answer on its components yet due to the fact that none of
the interpretations can fit all the available experiments
satisfactorily. The assignment of $X(3872)$ as a $2^3P_1$ charmonium
state, even without the mass gap problem \footnote{The mass of
$2^3P_1$ charmonium predicted by the quark model is about 100 MeV
larger than the measured $X(3872)$.} as claimed by calculations
based on the Lattice QCD recently in \cite{Y. Chen}, also bears
other difficulties. The tiny decay width of $X(3872)$, whose upper
bound is 2.3 MeV with 90\% confidence level, is much less than the
number predicted in theory \cite{swanson}. Another puzzle is the G
parity violation indicated by the measurement of the ratio of
branching fractions ${ BR(X \to J/\psi \pi^{+} \pi^{-} \pi^{0})\over
BR(X \to J/\psi \pi^{+} \pi^{-})}=1.0 \pm 0.4 \pm 0.3$ \cite{G
parity experiments 1,G parity experiments 2,G parity experiments 3}.
The difficulties of the charmonium interpretation invoke various
models for the structure of $X(3872)$, such as multi-quark
state\cite{multiquark state 1,multiquark state 2}, hybrid meson
\cite{hybrid meson}, nuclear-like molecular state \cite{molecular
1,molecular 2,molecular 3,molecular 4,molecular 5} and so on.  In
one word, the inner structure of $X(3872)$ is still not settled
down.

In addition to the intriguing particle $X(3872)$, other heavy hidden
charm mesons $X/Y(3940)$, $Z(3930)$ and $Y(4260)$ mentioned above
also attract comprehensive attention recently \cite{swanson}, among
which $Z(3930)$ can be well established as the first radial excited
states of tensor charmonium $\chi_{c2}$ reasonably and will be left
out in this paper.
 Even though  the experimental results of $h_c$
and $\eta_c^{\prime}$ are essentially consistent with theoretical
expectations, there are still some particular aspects deserving
further investigations \cite{Colangelo 1,Colangelo 2,hc in
pQCD,etac}. Besides, we also predict the production rate of
$h_c^{\prime}$ state in the weak $B_c$ decays, which has not
discovered. In addition, the $2S-1D$ mixing of $\psi(3686)$ and
$\psi(3770)$, which is of great interest in quarkonium physics, is
considered in the weak decays of $B_c$. More important, we also
investigate the production of $Y(4260)$ and $\psi(4415)$ in the weak
$B_c$ decays as the mixing of  $4S$ and $3D$ states. For the
completeness, the $3S-2D$ mixing of $\psi(4040)$ and $\psi(4160)$ in
the $B_c$ decay is also included.

In this work, we do not attempt to discuss all the explanations of
these states. Instead, we concentrate on the assignment of these
heavy mesons as charmonium states with the possible quantum numbers
constrained by the available experiments and  then study their
production properties in the $B_c$ decays. To be more specific, we
will assign the $X(3872)$ as a $2^3P_1$ charmonium, the $Y(4260)$ as
a $4^3S_1$ charmonium, the $X(3940)$ being either $3^1S_0$ or
$2^3P_1$ charmonium, and $2^3P_0 (c \bar{c})$ state for the
$Y(3940)$, the quantum numbers of which as the charmonium states are
most favored by the current experiments \cite{Colangelo 1,klempt},
although there is overpopulation of $2^3P_1 (c \bar{c})$ meson in
the charmonium family. The non-leptonic weak decays  $B^{+} \to
X(3872) K^{+}$ and $B_c \to X(3872) \pi (K)$ have been studied in
\cite{liu, wang} in order to pry information of the inner structure
of $X(3872)$. It is found there that different $J^{PC}$ assignment
of $X(3872)$ will give quite different decay rate for future
experiments to measure. Here we will focus on the semi-leptonic weak
production of charmonium particles in $B_c$ decays, where it is
theoretically easier compared with that of non-leptonic decays. We
will see that the different assignments of $J^{PC}$ quantum number
to $X(3943)$ will give remarkable different branching ratios of
semi-leptonic decays. Therefore our predictions can be used by
future experiments   to test the quark structures of these mesons.
The main job of calculating the branching fractions of the
semi-leptonic decays of $B_c$ is to properly evaluate the hadronic
matrix elements for $B_c \to M_{c \bar {c}}$ ($M$= pseudo-scalar
($P$), scalar ($S$), vector ($V$) or axial vector ($A$) charmonium),
namely the transition form factors.

The precise calculations of form factors are very complicated due to
the non-perturbative QCD effects in the hadron as a bound state.
Several methods have been developed to deal with this problem on the
market so far, such as simple quark model \cite{quark model},
light-front approach \cite{light front QCD 1,light front QCD 2,light
front QCD 3}, QCD sum rules (SVZ) \cite{QCDSR 1,QCDSR 2}, light-cone
QCD sum rules \cite{LCQCDSR 1,LCQCDSR 2,LCQCDSR 3}, perturbative QCD
factorization approach \cite{PQCD 1,PQCD 4}. Although the QCD sum
rules approach has made a big success, short distance expansion
fails in non-perturbative condensate when applying the three-point
sum rules  to the computations of form factors in the large momentum
transfer or large mass limit of heavy meson decays.
 The light-cone
QCD sum rules,  as a marriage of QCD sum rules techniques and the
theory of hard exclusive processes, were developed in an attempt to
overcome the difficulties \cite{Braun LCSR} involved in the SVZ sum
rules. The basic idea of light-cone QCD sum rules \cite{light front
QCD 1,light front QCD 2,light front QCD 3,Braun LCSR,perspective of
QCDSR} is to adopt the twist expansion of correlation functions near
the light-cone instead of the dimension expansion of operators at
short distance. Therefore, the essential inputs in the light-cone
QCD sum rules is the hadronic distribution amplitudes other than
vacuum condensates in the QCD sum rules. One important advantage of
light-cone QCD sum rules is that it allows a systematic inclusion of
both hard scattering effects and soft contributions
\cite{perspective of QCDSR}. In view of the above arguments, we will
estimate the form factors for $B_c$ to charmonium states based on
the light-cone QCD sum rules approach in this work.

The structure of this paper is organized as follows:  we  first
display the light-cone distribution amplitudes of various charmonium
states  in section \ref{distribution amplitudes}. The light-cone QCD
sum rules for the form factors responsible for the decay modes $B_c
\to M_{c \bar{c}}$ are derived in section \ref{Standard procedure}.
The numerical computations of form factors in light-cone QCD sum
rules are performed in  section \ref{Numerical results}. The decay
rates for semileptonic decays of $B_c$ to various charmonium states,
a brief analysis on comparisons with the results that obtained with
the help of other approaches in the literature and discussions on
the S-D mixing of $\psi(3686)$ and $\psi(3770)$  in the weak decay
of $B_c$ are also included in this section. The last section is
devoted to our
 conclusion.

\section{The light-cone distribution amplitudes   of charmonium states}
\label{distribution amplitudes}

The  light-cone distribution amplitudes (LCDAs) of pseudoscalar
charmonium can be defined by the following non-local matrix element
\cite{hsiang-nan li}
\begin{eqnarray}
\langle P(p)|\bar{c}(x)_{\alpha} c(0)_{\beta} | 0 \rangle = - {i
\over 4} f_{P} \int_0^1 du e^{i u p \cdot x} [(\gamma_5 \not
p)_{\beta \alpha} \phi^{v}(u) + m_{P} (\gamma_5)_{\beta \alpha}
\phi^{s}(u)], \label{pseudoscalar DAs}
\end{eqnarray}
where $\phi^{v}(u)$ and $\phi^{s}(u)$ are twist-2 and twist-3 LCDAs
of the pseudoscalar charmonium respectively. The decay constant
$f_{P}$ can be determined generally by decay width of the double
photons decay of the pseudoscalar meson as \cite{Bagchi}
\begin{eqnarray}
\Gamma(P \rightarrow \gamma \gamma)={4 (4 \pi \alpha)^2 f_P^2 \over
81 \pi m_P}.
\end{eqnarray}
Making use of the branching fractions of $\eta_c \to \gamma \gamma$
and the full width of $\eta_c$ \cite{PDG}
\begin{eqnarray}
{\rm{BR}}(\eta_c \to \gamma \gamma )=(2.8 \pm 0.9) \times 10^{-4},
\,\,\, \Gamma_{\eta_c}=(25.5\pm 3.4) {\rm{MeV}},
\end{eqnarray}
we can achieve the decay constant $f_{\eta_c}$ as $401^{+65}_{-76}
{\rm{MeV}}$. However, there is no data on $\eta_c^{\prime \prime}
\to \gamma \gamma$ till now, hence it is impossible to extract the
decay constant of $\eta_c^{\prime \prime}$ directly from the
experiments. In view of this point, we fix the  decay constant
$f_{\eta_c^{\prime \prime}}$ through the assumption,
${f_{\eta_c^{\prime \prime}} \over f_{\eta_c}}={f_{\psi^{\prime
\prime}} \over f_{J/\psi}}$, which has been used in \cite{K.T. Chao}
before. The decay constant of vector charmonium can be derived
through leptonic decay $V\rightarrow e^{+} e^{-}$ as
\begin{eqnarray}
f_{V}=\sqrt{{3 m_{V} \Gamma_{V \to ee} \over 4 \pi \alpha^2 Q_c^2}}.
\label{vector meson decay constants 1}
\end{eqnarray}
Combining the above relation and the data given in \cite{PDG}
\begin{eqnarray}
\Gamma_{J/\psi \rightarrow ee}=(5.55 \pm 0.14 \pm 0.02) {\rm{keV}},
\,\,\,\Gamma_{\psi^{\prime} \rightarrow ee}=(2.48 \pm
0.06){\rm{keV}}, \,\,\, \Gamma_{\psi^{\prime \prime} \rightarrow
ee}=(0.86 \pm 0.07){\rm{keV}},
\end{eqnarray}
we can obtain the decay constants of $\psi(nS) (n=1,2,3)$ as
\begin{eqnarray}
f_{J/\psi}=416^{+5}_{-6} {\rm{MeV}}, \,\,\,
f_{\psi^{\prime}}=304^{+3}_{-4} {\rm{MeV}}, \,\,\, f_{\psi ^{\prime
\prime}}=187 \pm 8 {\rm{MeV}}. \label{vector meson decay constants
2}
\end{eqnarray}
In light of   the assumption  mentioned above, we arrive at the
decay constant of $\eta_c^{\prime \prime}$ as $180^{+27}_{-32}
{\rm{MeV}}$. Moreover,  the decay constant of $\eta_c^{\prime}$ can
be   determined as  $293^{+48}_{-56} {\rm{MeV}}$.

It needs to be pointed out   that the tensor structure, which is
suppressed in the heavy quark limit, has been neglected in the right
hand side of the Eq. (\ref{pseudoscalar DAs}). When it comes to the
explicit forms of $\phi^{v}(u)$ and $\phi^{s}(u)$, we will adopt a
simple model advocated in \cite{chernyak}. Firstly,  one should
write down the Schr$\rm{\ddot{o}}$dinger equal-time wave function
$\Psi_{Sch}(r)$ for the Coulomb potential, and then perform the
Fourier transformation of it to the momentum space as
$\Psi_{Sch}(k)$. Next, in terms of the substitution assumption
proposed in \cite{terentev} (see also Eq. (\ref{substitution
assumption})), we can derive the expression of wave function
$\Psi_{Sch}(x_i, \mathbf{k}_{\perp})$ from $\Psi_{Sch}(k)$, where
the momentum fractions $x_1$, $x_2$ of $c$ and $\bar{c}$ quarks in
the charmonium satisfy the relation $x_1+x_2=1$. Finally, one can
achieve at the LCDAs of charmonium $\Psi_{Sch}(x_i)$ by integrating
over the transverse momentum $\mathbf{k}_{\perp}$. Based on this
prescription, we can obtain the LCDAs for $\eta_c^{\prime \prime}$
as
\begin{eqnarray}
\phi^{v}(x)&=&10.8 x(1-x)\bigg \{{x (1-x) [1-4x(1-x)(1+{v^2 \over
27})]^2 \over [1-4 x(1-x)(1-{v^2 \over 9})]^3} \bigg \}^{1-v^2},
\nonumber \\
\phi^{s}(x)&=&2.1 \bigg \{{x (1-x) [1-4x(1-x)(1+{v^2 \over 27})]^2
\over [1-4 x(1-x)(1-{v^2 \over 9})]^3} \bigg \}^{1-v^2},
\end{eqnarray}
where the variable $v$ reflects the mean charm quark velocity and is
taken as $v^2=0.30 \pm 0.05$ \cite{chernyak} in the numerical
analysis.  To be more clear, the shape of the distribution amplitude
$\phi^{v}(x)$ is shown in Fig. \ref{distribution amplitudes of 3S}
with $v^2=0.3$.

\begin{figure}[tb]
\begin{center}
\begin{tabular}{ccc}
\includegraphics[scale=0.8]{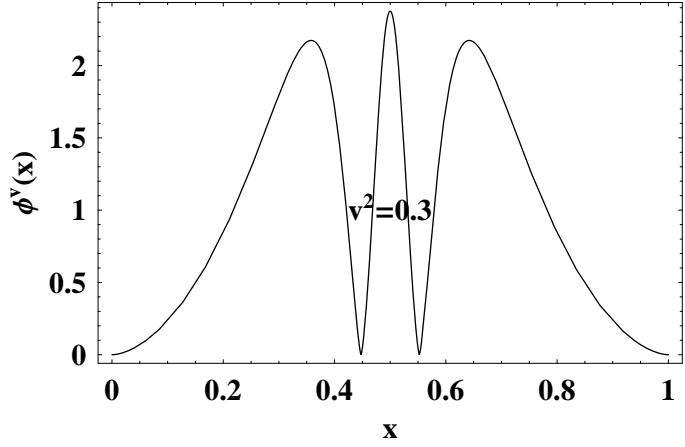}
\vspace{-2 cm}
\end{tabular}
\caption{The shape of the distribution amplitude $\phi^{v}(x)$ for
$\eta_c^{\prime \prime}$ with $v^2=0.3$.} \label{distribution
amplitudes of 3S}
\end{center}
\end{figure}

Similarly, we can also derive the LCDAs for $\eta_c^{\prime}$
\begin{eqnarray}
\phi^{v}(x)&=&10.6 x(1-x)\bigg \{{x (1-x) (1-2x)^2 \over [1-4
x(1-x)(1-{v^2 \over 4})]^2} \bigg \}^{1-v^2},
\nonumber \\
\phi^{s}(x)&=&2.1 \bigg \{{x (1-x) (1-2x)^2 \over [1-4 x(1-x)(1-{v^2
\over 4})]^2} \bigg \}^{1-v^2}.
\end{eqnarray}

Similarly, the LCDAs of scalar charmonium state can be defined by
\begin{eqnarray}
\langle S(p)|\bar{c}(x)_{\alpha} c(0)_{\beta} | 0 \rangle = {1 \over
4} f_{S} \int_0^1 du e^{i u p \cdot x} [(\not p)_{\beta \alpha}
\chi^{v}(u) + m_{S} (I)_{\beta \alpha} \chi^{s}(u)], \label{scalar
DAs}
\end{eqnarray}
with $\chi^{v}(u)$ and $\chi^{s}(u)$ being the twist-2 and twist-3
DAs for the scalar meson respectively. Based on the method of
building the model for heavy quarkonium's distribution amplitudes
given above, we can obtain the explicit forms of distribution
amplitudes as
\begin{eqnarray}
\chi^{v}(x)&=&90.2 x(1-x)(1-2x)\bigg \{{x (1-x) (1-2x)^4  \over [1-4
x(1-x)(1-{v^2 \over 9})]^3} \bigg \}^{1-v^2},
\nonumber \\
\chi^{s}(x)&=&1.9 \bigg \{{x (1-x) (1-2x)^4  \over [1-4
x(1-x)(1-{v^2 \over 9})]^3} \bigg \}^{1-v^2},
\end{eqnarray}
for the $2^3P_0$ charmonium $\chi_{c0}^{\prime}$. In addition, the
decay constant of $\chi_{c0}^{\prime}$ can be calculated as $263
^{+6}_{-7} \rm{MeV}$ making use of the assumption
${f_{\chi_{c0}^{\prime}} \over f_{\chi_{c0}}}={f_{\psi^{\prime}}
\over f_{J/\psi}}$  and the value of $f_{\chi_{c0}}$, which was
estimated to be $360 \rm{MeV}$ in Ref. \cite{SVZ rep,hsiang-nan li}.

The non-local matrix element associating with the vector charmonium
can be decomposed as \cite{chernyak}
\begin{eqnarray}
\langle V(p, \epsilon)|\bar{c}(x)_{\alpha} c(0)_{\beta} | 0 \rangle
&=& {1 \over 4} \int_0^1 du e^{i u p \cdot x} \bigg\{ f_{V} m_V
(\not \epsilon^{\ast}-{\epsilon^{\ast} \cdot x\over p \cdot x} \not
p) V_{\perp} (u) +f_{V} m_V {\epsilon^{\ast} \cdot x\over p \cdot x}
\not p V_{L}(u) \nonumber \\
&&+ f_{V}^{T} \not \epsilon \not p V_{T}(u)+ {1 \over 4}(f_V- {2 m_c
\over m_V} f_V^T) \epsilon_{\mu \nu \alpha \beta} \gamma^{\mu}
\gamma^{5} \epsilon^{\ast \nu} p^{\alpha} x^{\beta}
V_{A}(u)\bigg\}_{\beta \alpha},
\end{eqnarray}
where $V_{L}(u)$,  $V_{T}(u)$ are the leading twist longitudinal and
transverse LCDAs of vector charmonium, and $V_{\perp} (u)$,
$V_{A}(u)$ are the twist-3 ones. Following the methods described
above, we can deduce the manifest expressions of these distribution
amplitudes as
\begin{eqnarray}
V_{L}(x)&=&10.8 x(1-x)\bigg \{{x (1-x) ((1-2x)^2 [1-4x(1-x)(1+{v^2
\over 16})]^2 \over [1-4 x(1-x)(1-{v^2 \over 16})]^4} \bigg
\}^{1-v^2},
\nonumber \\
V_{\perp}(x)&=&1.74 [1+(1-2x)^2]\bigg \{{x (1-x) ((1-2x)^2
[1-4x(1-x)(1+{v^2 \over 16})]^2 \over [1-4 x(1-x)(1-{v^2 \over
16})]^4} \bigg \}^{1-v^2}
\nonumber \\
V_{T}(x)&=&V_{A}(x)=V_{L}(x),
\end{eqnarray}
for the $4^3S_1$ charmonium. In the numerical calculations, the
decay constants $f_{V}$ and $f_{V}^{T}$ are assumed to be equal
\cite{hsiang-nan li}. The purely leptonic decay of $Y(4260)
\rightarrow e^{+} e^{-}$ ie estimated to be $0.72 {\rm{keV}}$
\cite{klempt}, from which we can obtain the decay constant
$f_{Y(4260)}$ as $176 \rm{MeV}$.

Similarly, the light-cone distribution amplitudes of $3^3D_1$
charmonium can be derived as
\begin{eqnarray}
V_{L}(x)&=&2.8 x(1-x)\bigg \{{x^2 (1-x)^2 (1-2x)^6 \over [1-4
x(1-x)(1-{v^2 \over 25})]^5} \bigg \}^{1-v^2},
\nonumber \\
V_{\perp}(x)&=&0.62 [1+(1-2x)^2]\bigg \{{x^2 (1-x)^2 (1-2x)^6 \over
[1-4 x(1-x)(1-{v^2 \over 25})]^5} \bigg \}^{1-v^2},
\nonumber \\
V_{T}(x)&=&V_{A}(x)=V_{L}(x).
\end{eqnarray}
With the hypothesis ${f_{\psi(3^3D_1)} \over
f_{\psi(1^3D_1)}}={f_{\psi^{\prime \prime}} \over f_{J/\psi}}$ and
$f_{\psi(1^3D_1)}=47.8 \,\, \rm{MeV}$ \cite{Rosner}, we can achieve
the value of $f_{\psi(3^3D_1)}$ as $21.5 ^{+0.6}_{-0.5} {\rm{MeV}}$
under the above assumption. This is a quite small decay constant,
comparing with that of the corresponding S-wave charmonium states.
This will surely lead to the quite small form factors, since the
transition form factors are proportion to the decay constant of the
final state meson as can be observed form the light-cone sum rules
in the next section. In the same way, we can arrive at the decay
constant of $\psi(2^3D_1)$ as $f_{\psi(2^3D_1)}=34.9 ^{+0.8}_{-0.9}
{\rm{MeV}}$.

For the sake of investigating the $2S-1D$ mixing of $\psi(3686)$ and
$\psi(3770)$, it is necessary to derive the light-cone distribute
amplitudes for the $\psi(2^3S_1)$ and $\psi(1^3D_1)$ based on the
model discussed above. To be more specific, the LCDAs for
$\psi(2^3S_1)$ can be given by
\begin{eqnarray}
V_{L}(x)&=&V_{T}(x)=V_{A}(x)=10.6 x(1-x)\bigg \{{x (1-x) (1-2x)^2
\over [1-4 x(1-x)(1-{v^2 \over 4})]^2} \bigg \}^{1-v^2},
\nonumber \\
V_{\perp}(x)&=& 1.7[1+(1-2x)^2]\bigg \{{x (1-x) (1-2x)^2 \over [1-4
x(1-x)(1-{v^2 \over 4})]^2} \bigg \}^{1-v^2};
\end{eqnarray}
while the LCDAs for $\psi(1^3D_1)$ can read as
\begin{eqnarray}
V_{L}(x)&=&V_{T}(x)=V_{A}(x)=3.6 x(1-x)\bigg \{{x^2 (1-x)^2 (1-2x)^2
\over [1-4 x(1-x)(1-{v^2 \over 9})]^3} \bigg \}^{1-v^2},
\nonumber \\
V_{\perp}(x)&=& 0.77[1+(1-2x)^2]\bigg \{{x^2 (1-x)^2 (1-2x)^2 \over
[1-4 x(1-x)(1-{v^2 \over 9})]^3} \bigg \}^{1-v^2}.
\end{eqnarray}

Moreover, we also would like to present the explicit forms of LCDAs
for $\psi(3^3S_1)$ and $\psi(2^3D_1)$, which are essential to study
the $3S-2D$ mixing of $\psi(4040)$ and $\psi(4160)$. The LCDAs for
$\psi(3^3S_1)$ can be calculated as
\begin{eqnarray}
V_{L}(x)&=&V_{T}(x)=V_{A}(x)=10.8 x(1-x)\bigg \{{x (1-x)
[1-4x(1-x)(1+{v^2 \over 27})]^2 \over [1-4 x(1-x)(1-{v^2 \over
9})]^3} \bigg \}^{1-v^2},
\nonumber \\
V_{\perp}(x)&=&1.7 [1+(1-2x)^2]\bigg \{{x (1-x) [1-4x(1-x)(1+{v^2
\over 27})]^2 \over [1-4 x(1-x)(1-{v^2 \over 9})]^3} \bigg
\}^{1-v^2};
\end{eqnarray}
while the LCDAs for $\psi(2^3D_1)$ are given by
\begin{eqnarray}
V_{L}(x)&=&V_{T}(x)=V_{A}(x)=3.2 x(1-x)\bigg \{{x^2 (1-x)^2 (1-2x)^4
\over [1-4 x(1-x)(1-{v^2 \over 16})]^4} \bigg \}^{1-v^2},
\nonumber \\
V_{\perp}(x)&=&0.70[1+(1-2x)^2]\bigg \{{x^2 (1-x)^2 (1-2x)^4 \over
[1-4 x(1-x)(1-{v^2 \over 16})]^4} \bigg \}^{1-v^2}.
\end{eqnarray}

As far as the axial-vector charmonium is concerned, the
corresponding non-local matrix element can be analyzed as \cite{k.c
yang}
\begin{eqnarray}
\langle A(p, \epsilon)|\bar{c}(x)_{\alpha} c(0)_{\beta} | 0 \rangle
&=& -{i \over 4} \int_0^1 du e^{i u p \cdot x} \bigg\{ f_{A} m_A
(\not \epsilon^{\ast}-{\epsilon^{\ast} \cdot x\over p \cdot x} \not
p) \gamma_5 g_{\perp}^{(a)} (u) +f_{A} m_A {\epsilon^{\ast} \cdot
x\over p \cdot x}
\not p \gamma_5 \phi_{\parallel}(u) \nonumber \\
&&+ f_{A}^{T} \not \epsilon \not p \gamma_5 \phi_{\perp}(u)+ {1
\over 4}(f_A- {2 m_c \over m_A} f_A^T) \epsilon_{\mu \nu \alpha
\beta} \gamma^{\mu} \epsilon^{\ast \nu} p^{\alpha} x^{\beta}
g_{\perp}^{v}(u)\bigg\}_{\beta \alpha},
\end{eqnarray}
where $\phi_{\parallel}(u)$, $\phi_{\perp}(u)$ are of twist-2, and
$g_{\perp}^{v}(u)$ and $g_{\perp}^{a}(u)$ are the twist-3 LCDAs of
axial-vector charmonium. As for the $n ^{3}P_1$ states,
$\phi_{\parallel}(u)$, $g_{\perp}^{v}(u)$ and $g_{\perp}^{a}(u)$ are
symmetric under the exchange of momentum fractions $u$ and $1-u$,
but $\phi_{\perp}(u)$ is anti-symmetric under this exchange. On the
contrary, $\phi_{\perp}(u)$ is symmetric for $n ^{1}P_1$ states,
while $\phi_{\parallel}(u)$, $g_{\perp}^{v}(u)$ and
$g_{\perp}^{a}(u)$ are anti-symmetric in this case. Following the
procedure of constructing the wave functions for heavy quarkonium
shown above, we can arrive at
\begin{eqnarray}
\phi_{\perp}(x)&=&90.2 x(1-x)(1-2x)\bigg \{{x (1-x) (1-2x)^4  \over
[1-4 x(1-x)(1-{v^2 \over 9})]^3} \bigg \}^{1-v^2},
\end{eqnarray}
for the $2 ^{3}P_1$ charmonium,
\begin{eqnarray}
\phi_{\perp}(x)&=&10.6 x(1-x)\bigg \{{x (1-x) (1-2x)^2  \over [1-4
x(1-x)(1-{v^2 \over 4})]^2} \bigg \}^{1-v^2},
\end{eqnarray}
for the $1 ^{1}P_1$ charmonium $h_c$, and
\begin{eqnarray}
\phi_{\perp}(x)&=&9.5 x(1-x)\bigg \{{x (1-x)(1-2x)^4 \over [1-4
x(1-x)(1-{v^2 \over 9})]^3} \bigg \}^{1-v^2},
\end{eqnarray}
for the $2 ^{1}P_1$ charmonium $h_c^{\prime}$.

As can be seen below, only the leading twist LCDA of axial-vector
charmonium $\phi_{\perp}(u)$ is involved in the light-cone QCD sum
rules of form factors, hence, the
  expressions of other three distribution amplitudes will not be
shown here. Notice that we       assume the decay constants $f_{A}
=f_{A}^{T}$ in the practical numerical analysis,   the same as that
for the vector charmonium. Meanwhile, the decay constant of $P$ wave
charmonium $2^3P_1 (c \bar{c})$,   was estimated to be $207
{\rm{MeV}}$  \cite{g.l wang} very recently. Moreover, the decay
constant of $h_c$  and $h_c^{\prime}$ are taken the same as that for
the $\chi_{c1}$ in \cite{suzuki} and $\chi_{c1}^{\prime}$
respectively, namely $f_{h_c} =f_{\chi_{c1}}=335 {\rm{MeV}}$
\cite{hsiang-nan li}, $f_{h_c^{\prime}} =f_{\chi_{c1}^{\prime}}=207
{\rm{MeV}}$.

\section{Light-cone QCD sum rules for the weak transition form factors}

\label{Standard procedure}

For the semi-leptonic decays of $B_c \to M_{c \bar{c}}l
\bar{\nu}_l$, the effective weak Hamiltonian is given by
\begin{eqnarray}
 \mathcal{H}_{eff}(b \to c l\bar \nu_l)={G_{F} \over \sqrt{2}}V_{cb}
 \bar{c}\gamma_{\mu}(1-\gamma_5)b \,
 \bar{l}\gamma^{\mu}(1-\gamma_5)\nu_l +h.c. \, ,
\end{eqnarray}
where $V_{cb}$ is the corresponding Cabbibo-Kobayashi-Maskawa (CKM)
matrix element. In order to estimate the decay rates of $B_c \to
M_{c \bar{c}}l \bar{\nu}_l$, we need to calculate the hadronic
matrix element $\langle
M_{c\bar{c}}|\bar{c}\gamma_{\mu}(1-\gamma_5)b| B_c \rangle$ at
first, which can be conventionally parameterized in the following
forms:
\begin{eqnarray}
\langle P_{c\bar{c}}(p)|\bar{c}\gamma_{\mu}b| B_c(p+q) \rangle &=&
f_{+}(q^2) p_{\mu} +f_{-}(q^2) q_{\mu},
\\
\langle S_{c\bar{c}}(p)|\bar{c}\gamma_{\mu}\gamma_5 b| B_c(p+q)
\rangle &=& -i[f_{+}(q^2) p_{\mu} +f_{-}(q^2) q_{\mu}],
\\
\langle V_{c\bar{c}}(p)|\bar{c}\gamma_{\mu}(1-\gamma_5)b| B_c(p+q)
\rangle &=& {2 V(q^2) \over m_{B_c} +m_V} \epsilon_{\mu \alpha \beta
\gamma} \epsilon^{\ast \alpha} q^{\beta} p^{\gamma} -i (m_{B_c}
+m_V) A_1(q^2) \epsilon^{\ast}_{\mu} \nonumber \\
&&+ {i A_2(q^2) \over m_{B_c} +m_V} (\epsilon^{\ast} \cdot q)
(2p+q)_{\mu} + {i A_3(q^2) \over m_{B_c} +m_V} (\epsilon^{\ast}
\cdot q) q_{\mu},
\\
\langle A_{c\bar{c}}(p)|\bar{c}\gamma_{\mu}(1-\gamma_5)b| B_c(p+q)
\rangle &=& -{2i A(q^2) \over m_{B_c}-m_A} \epsilon_{\mu \alpha
\beta \gamma} \epsilon^{\ast \alpha} q^{\beta} p^{\gamma} -
(m_{B_c}-m_A) V_1(q^2) \epsilon^{\ast}_{\mu} \nonumber \\
&& + { V_2(q^2) \over m_{B_c}-m_A} (\epsilon^{\ast} \cdot q)
(2p+q)_{\mu} + { V_3(q^2) \over m_{B_c}-m_A} (\epsilon^{\ast} \cdot
q) q_{\mu},
\end{eqnarray}
where  the   anti-symmetric forth rank tensor is defined as
${\rm{Tr}}[\gamma_{\mu} \gamma_{\nu} \gamma_{\rho} \gamma_{\sigma}
\gamma_5]=4 i \epsilon_{\mu \nu \rho \sigma}$.

Below, we will derive the general formulae for the form factors of
$B_c \to M_{c \bar{c}} (M=P, S, V, A)$ in the light-cone QCD sum
rules approach. Following  Ref. \cite{chiral current 1,chiral
current 2}, the correlation function is selected with the insertion
of chiral current, to which the twist-3 distribution amplitudes of
final states do not contribute at all for semi-leptonic $B_{c} \to
P(S)$ decays. As for the $B_{c} \to V(A)$ decays, the two-particle
distribution amplitudes of twist-3 also have no effect on the
correlation function with the insertion of chiral current in the
heavy charm quark mass limit. Besides, the twist-3 distribution
amplitudes relating to  the three-particle $\bar{c} c g$ Fock state,
which are suppressed by a factor $(\Lambda_{QCD}/m_{\bar{c} c})^2$
\cite{chernyak} with $m_{\bar{c} c}$ being the mass of the
charmonium, are also omitted in this work. The estimation of
correlation functions in the QCD representation can be carried out
following the standard prescription given in \cite{LCSR method
1,LCSR method 2}.

\subsection{Light-cone QCD sum rules for the weak transition form
factors of  $B_c \to P_{c \bar{c}}$}

Based on the above analysis, we firstly construct the following
correlator $\Pi_{\mu}(p,q)$ with the insertion of the chiral
current:
\begin{eqnarray}
\Pi_{\mu}(p,q)=i \int d^4 x e^{i q \cdot x} \langle P_{c
\bar{c}}(p)|T\{\bar{c}(x)\gamma_{\mu}(1+\gamma_5) b(x), \bar{b}(0) i
(1+\gamma_5) c(0)\}| 0\rangle. \label{correlator of P meson}
\end{eqnarray}
One character of this correlation function is that twist-3
distribution amplitude of pseudoscalar charmonium has no influence
on it and therefore the theoretical uncertainties can reduced
considerably in this way. Inserting  the complete sets of hadronic
states with the quantum numbers the same as $B_c$ and making use of
the following definition
\begin{eqnarray}
\langle B_c |\bar{b} i (1+\gamma_5)c | 0 \rangle ={m_{B_c}^2 f_{B_c}
\over m_b+m_c},
\end{eqnarray}
we can arrive at the hadronic representation of correlation function
(\ref{correlator of P meson}) as below:
\begin{eqnarray}
\Pi_{\mu}(p,q)&=& {\langle
P_{c\bar{c}}(p)|\bar{c}\gamma_{\mu}(1+\gamma_5)b| B_c(p+q) \rangle
\langle B_c (p+q) |\bar{b} i (1+\gamma_5)c | 0 \rangle \over
m_{B_c}^2 -(p+q)^2}
\nonumber \\
&&+ \sum_{h} {\langle
P_{c\bar{c}}(p)|\bar{c}\gamma_{\mu}(1+\gamma_5)b| h(p+q) \rangle
\langle h (p+q) |\bar{b} i (1+\gamma_5)c | 0 \rangle \over
m_{h}^2 -(p+q)^2} \nonumber \\
&=& {m_{B_c}^2 f_{B_c} (f_{+}(q^2) p_{\mu} +f_{-}(q^2) q_{\mu})
\over (m_b+m_c)(m_{B_c}^2 -(p+q)^2)}+ \int_{s_0^{B_c}}^{\infty} ds {
\rho^h_{+}(s,q^2) p_{\mu} +\rho^h_{-}(s,q^2) q_{\mu} \over
s-(p+q)^2} ,
\end{eqnarray}
where we have expressed the contributions from higher states of the
$B_c$ channel in the form of dispersion integral with $s_0^{B_c}$
being the threshold parameter corresponding to the $B_c$ channel. On
the other hand, we can also calculate the correlation function at
the quark level:
\begin{eqnarray}
\Pi_{\mu}(p,q)&=& \Pi_{+}^{QCD}(q^2,(p+q)^2) p_{\mu}
+\Pi_{-}^{QCD}(q^2,(p+q)^2) q_{\mu} \nonumber \\
&=&\int_{(m_b+m_c)^2}^{\infty} ds {1 \over \pi} {{\rm{Im}}\,
\Pi_{+}^{QCD}(s,q^2) \over s-(p+q)^2} p_{\mu}
+\int_{(m_b+m_c)^2}^{\infty} ds {1 \over \pi} { {\rm{Im}} \,
\Pi_{-}^{QCD}(s,q^2) \over s-(p+q)^2} q_{\mu}.
\end{eqnarray}
Utilizing the quark-hadron duality assumption
\begin{eqnarray}
\rho^h_{i}(s,q^2)={1 \over \pi}  {\rm{Im}}\, \Pi_{i}^{QCD}(s,q^2)
\Theta (s-s_0^h),
\end{eqnarray}
with $i=``+,-"$ and performing the Borel transformation
\begin{eqnarray}
\hat{\mathcal{B}}_{M^2}=\lim_{\stackrel{-(p+q)^2,n \to
\infty}{-(p+q)^2/n=M^2}} \frac{(-(p+q)^2)^{(n+1)}}{n!}\left(
\frac{d}{d(p+q)^2}\right)^n,
\end{eqnarray}
with variable $(p+q)^2$ to both two representations of the
correlation function, we can finally derive the sum rules for the
form factors
\begin{eqnarray}
f_i(q^2)={m_b+m_c \over \pi f_{B_c}m_{B_c}^2}
\int_{(m_b+m_c)^2}^{s_0^{B_c}} {\rm{Im}} \, \Pi_{i}^{QCD}(s,q^2)
{\rm{exp}}\bigg({m_{B_c}^2-s \over M^2}\bigg) ds. \label{formal sum
rules for P}
\end{eqnarray}

The QCD representation of correlation function (\ref{correlator of P
meson}) can be calculated in terms of operator product expansion
(OPE) in both of the large space-like region $(p+q)^2 \ll
-(m_b+m_c)^2$  and the   low momentum transfer region \cite{LCSR
method 2, B to D in LCSR} $q^2\leq (m_b-m_c)^2- 2 \Lambda_{QCD}
(m_b-m_c)\simeq 8.2 {\rm{GeV}^2}$, where the value of
$\Lambda_{QCD}$ is usually taken as 0.5 GeV. It is expected the
light-cone QCD sum rules approach for the transition form factors
will break down at large momentum transfer \cite{LCSR method 2},
since the light-cone expansion for the description of final state
meson is not well-pleasing in this case and the contributions from
the higher twists would be important. The leading order contribution
in the OPE can be gained simply by contracting the b-quark operators
in the correlator (\ref{correlator of P meson}) to a free b-quark
propagator
\begin{eqnarray}
\langle 0 |b(x) \bar{b}(0) | 0\rangle=\int {d^4k \over i (2 \pi)^4}
e^{-i k \cdot x} {\not{k} + m_b \over m_b^2 -k^2 },
\end{eqnarray}
which can be  represented by Fig.~\ref{correlation function at quark
level} intuitively.

\begin{figure}[tb]
\begin{center}
\begin{tabular}{ccc}
\includegraphics[scale=0.6]{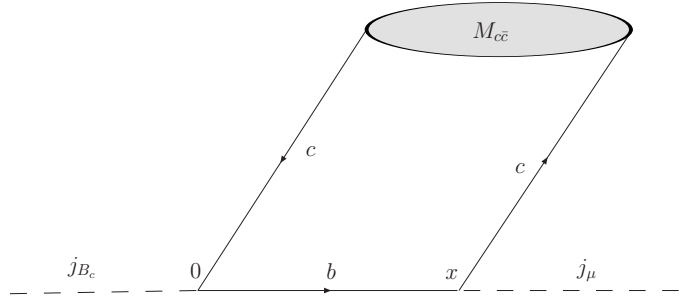}
\vspace{0 cm}
\end{tabular}
\caption{The tree level contribution to the correlation function Eq.
(\ref{correlator of P meson}), where the current $j_{B_c}(0)$
describe the $B_c$ channel and the current $j_{\mu}(x)$ is associate
with the $b \to c$ transition.} \label{correlation function at quark
level}
\end{center}
\end{figure}

Then we arrive at the correlation function at the quark level as
\begin{eqnarray}
\Pi_{+}(q^2,(p+q)^2)&=&-2 m_b f_{P_{c \bar {c}}} \int_0^1 du
{\phi^v(u) \over (q+up)^2 -m_b^2 +i \epsilon} +
\mathrm{contributions}\,\,\, \mathrm{from}\,\,\,
\mathrm{higher}\,\,\, \mathrm{twists}, \nonumber \\
\Pi_{-}(q^2,(p+q)^2)&=& 0 + \mathrm{contributions}\,\,\,
\mathrm{from}\,\,\, \mathrm{higher}\,\,\, \mathrm{twists},
\label{QCD representation}
\end{eqnarray}
where the higher twists contributions are at least from twist-4
distribution amplitudes of the pseudoscalar charmonium \cite{chiral
current 2,B to D in LCSR,fen zuo}. Substituting the Eq.(\ref{QCD
representation}) to Eq.(\ref{formal sum rules for P}), we can
finally derive the light-cone QCD sum rules for the form factors
$f_i(q^2)$  as below
\begin{eqnarray}
f_{+}(q^2)&=&{2 m_b (m_b+m_c) f_{P_{c \bar{c}}} \over f_{B_c}
m_{B_c}^2} {\rm{exp}} \bigg({m_{B_c}^2 \over M^2}\bigg)
\int_{\Delta}^1 {du \over u} \phi^{v}(u)  {\rm{exp}} \bigg[
-{m_b^2-\bar{u} (q^2-u p^2)\over u M^2}\bigg], \nonumber \\
f_{-}(q^2)&=&0,
\end{eqnarray}
up to the accuracy of twist-3 LCDAs, with
\begin{eqnarray}
\Delta={-(s-q^2-p^2)+\sqrt{(s-q^2-p^2)^2+4 p^2(m_b^2-q^2)}\over 2
p^2 }, \label{lower limit}
\end{eqnarray}
$p^2$ being the mass square for the corresponding charmonium state
($m_P^2$ in this subsection) and $s$ being the threshold value of
$B_c$ channel. It needs to be emphasized that the vanishing of
$f_{-}(q^2)$ up to the twist-3 LCDAs of pseudoscalar charmonium and
leading order of the strong coupling constant $\alpha_s$ is the
consequence of the large-recoil symmetry \cite{charles}, which
emerges in the case of large recoil momentum for the final state
meson and can be broken by the hard gluon corrections \cite{beneke}.

\subsection{Light-cone QCD sum rules for the weak transition form
factors of  $B_c \to S_{c \bar{c}}$}

Following the derivation of the light-cone sum rules for $B_c \to
P_{c \bar{c}}$, the correlation function of $B_c \to S_{c \bar{c}}$
can be written as
\begin{eqnarray}
\Pi_{\mu}(p,q)=i \int d^4 x e^{i q \cdot x} \langle S_{c
\bar{c}}(p)|T\{\bar{c}(x)\gamma_{\mu}(1+\gamma_5) b(x), \bar{b}(0) i
(1+\gamma_5) c(0)\}| 0\rangle. \label{correlator of S meson}
\end{eqnarray}
Matching the results of the above correlator calculated in the quark
level and hadron representation respectively and performing Borel
transformation with the variable $(p+q)^2$, we can achieve the
light-cone sum rules for the transition form factors as below
\begin{eqnarray}
f_{+}(q^2)&=&-{2 m_b (m_b+m_c) f_{S_{c \bar{c}}} \over f_{B_c}
m_{B_c}^2} {\rm{exp}} \bigg({m_{B_c}^2 \over M^2}\bigg)
\int_{\Delta}^1 {du \over u} \chi^{v}(u)  {\rm{exp}} \bigg[
-{m_b^2-\bar{u} (q^2-u p^2)\over u M^2}\bigg], \nonumber \\
f_{-}(q^2)&=&0, \label{sum rules of S meson}
\end{eqnarray}
where the lower limit of the integral $\Delta$ has been given in
Eq.(\ref{lower limit}).

\subsection{Light-cone QCD sum rules for the weak transition form
factors of  $B_c \to V_{c \bar{c}}$}

In the same way, the correlation function with the insertion of
chiral current for $B_c \to V_{c \bar{c}}$ can be chosen as
\begin{eqnarray}
\Pi_{\mu}(p,q)=i \int d^4 x e^{i q \cdot x} \langle V_{c
\bar{c}}(p)|T\{\bar{c}(x)\gamma_{\mu}(1-\gamma_5) b(x), \bar{b}(0) i
(1+\gamma_5) c(0)\}| 0\rangle. \label{correlator of V meson}
\end{eqnarray}
The hadronic representation of this correlator can be derived as
\begin{eqnarray}
\Pi_{\mu}(p,q)&=& {\langle
V_{c\bar{c}}(p)|\bar{c}\gamma_{\mu}(1-\gamma_5)b| B_c(p+q) \rangle
\langle B_c (p+q) |\bar{b} i (1+\gamma_5)c | 0 \rangle \over
m_{B_c}^2 -(p+q)^2}
\nonumber \\
&&+ \sum_{h} {\langle
V_{c\bar{c}}(p)|\bar{c}\gamma_{\mu}(1-\gamma_5)b| h(p+q) \rangle
\langle h (p+q) |\bar{b} i (1+\gamma_5)c | 0 \rangle \over
m_{h}^2 -(p+q)^2} \nonumber \\
&=& {2 m_{B_c}^2 f_{B_c} V(q^2)\over (m_b+m_c)(m_{B_c}+m_{V_{c
\bar{c}}})(m_{B_c}^2 -(p+q)^2)}\epsilon_{\mu \alpha \beta
\gamma}\epsilon^{\ast \alpha} q^{\beta} p^{\gamma} - i {m_{B_c}^2
f_{B_c}(m_{B_c}+m_{V_{c \bar{c}}}) A_1(q^2)\over
(m_b+m_c)(m_{B_c}^2 -(p+q)^2)} \epsilon_{\mu}^{\ast} \nonumber \\
&&+ i { m_{B_c}^2 f_{B_c} (\epsilon^{\ast} \cdot q) [A_2(q^2)
(2p+q)_{\mu}+A_3(q^2) q_{\mu}]\over (m_b+m_c)(m_{B_c}+m_{V_{c
\bar{c}}})(m_{B_c}^2 -(p+q)^2)} + \int_{s_0^{B_c}}^{\infty} ds {
\rho^h_{V}(s,q^2) \over s-(p+q)^2} \epsilon_{\mu \alpha \beta
\gamma}\epsilon^{\ast \alpha} q^{\beta} p^{\gamma}\nonumber \\
&&+ \int_{s_0^{B_c}}^{\infty} ds { \rho^h_{A_1}(s,q^2) \over
s-(p+q)^2} \epsilon_{\mu}^{\ast} + \int_{s_0^{B_c}}^{\infty} ds {
\rho^h_{A_2}(s,q^2) (2p+q)_{\mu}+\rho^h_{A_3}(s,q^2) q_{\mu}\over
s-(p+q)^2} (\epsilon^{\ast} \cdot q).
\end{eqnarray}
Besides, the correlation function in Eq.(\ref{correlator of V
meson}) can also be formulated  as
\begin{eqnarray}
\Pi_{\mu}(p,q)&=& \Pi_{V}^{QCD}(q^2,(p+q)^2) \epsilon_{\mu \alpha
\beta \gamma}\epsilon^{\ast \alpha} q^{\beta} p^{\gamma}
-i \Pi_{A_1}^{QCD}(q^2,(p+q)^2) \epsilon_{\mu}^{\ast}\nonumber \\
&& + i \Pi_{A_2}^{QCD}(q^2,(p+q)^2) (\epsilon^{\ast} \cdot q) (2
p+q)_{\mu}+i \Pi_{A_3}^{QCD}(q^2,(p+q)^2) (\epsilon^{\ast} \cdot q) q_{\mu}
\nonumber \\
&=&\int_{(m_b+m_c)^2}^{\infty} ds {1 \over \pi}  { {\rm{Im}}\,
\Pi_{V}^{QCD}(s,q^2) \over s-(p+q)^2} \epsilon_{\mu \alpha \beta
\gamma}\epsilon^{\ast \alpha} q^{\beta} p^{\gamma} -i
\int_{(m_b+m_c)^2}^{\infty} ds {1 \over \pi}  { {\rm{Im}}\,
\Pi_{A_1}^{QCD}(s,q^2) \over s-(p+q)^2} \epsilon_{\mu}^{\ast} \nonumber \\
&&+ i \int_{(m_b+m_c)^2}^{\infty} ds {1 \over \pi}  { {\rm{Im}}\,
\Pi_{A_2}^{QCD}(s,q^2) (2p+q)_{\mu}+  {\rm{Im}}\,
\Pi_{A_3}^{QCD}(s,q^2) q_{\mu}\over s-(p+q)^2} (\epsilon^{\ast}
\cdot q) .
\end{eqnarray}
Matching these two representations of the correlator  and performing
the Borel transforming with the variable $(p+q)^2$ on them,   we can
obtain the light-cone QCD sum rules for the form factors of $B_c \to
V_{c \bar{c}}$ as
\begin{eqnarray}
V(q^2)&=&{(m_b+m_c)(m_{B_c}+m_{V_{c \bar{c}}}) \over 2 \pi
f_{B_c}m_{B_c}^2} \int_{(m_b+m_c)^2}^{s_0^{B_c}} {\rm{Im}} \,
\Pi_{V}^{QCD}(s,q^2) {\rm{exp}}\bigg({m_{B_c}^2-s \over M^2}\bigg)
ds, \nonumber \\
A_1(q^2)&=&{m_b+m_c \over  \pi f_{B_c}m_{B_c}^2 (m_{B_c}+m_{V_{c
\bar{c}}})} \int_{(m_b+m_c)^2}^{s_0^{B_c}} {\rm{Im}} \,
\Pi_{A_1}^{QCD}(s,q^2) {\rm{exp}}\bigg({m_{B_c}^2-s \over M^2}\bigg)
ds, \nonumber \\
A_2(q^2)&=&{(m_b+m_c)(m_{B_c}+m_{V_{c \bar{c}}}) \over  \pi
f_{B_c}m_{B_c}^2} \int_{(m_b+m_c)^2}^{s_0^{B_c}} {\rm{Im}} \,
\Pi_{A_2}^{QCD}(s,q^2) {\rm{exp}}\bigg({m_{B_c}^2-s \over M^2}\bigg)
ds, \nonumber \\
A_3(q^2)&=&{(m_b+m_c)(m_{B_c}+m_{V_{c \bar{c}}}) \over  \pi
f_{B_c}m_{B_c}^2} \int_{(m_b+m_c)^2}^{s_0^{B_c}} {\rm{Im}} \,
\Pi_{A_3}^{QCD}(s,q^2) {\rm{exp}}\bigg({m_{B_c}^2-s \over M^2}\bigg)
ds. \label{formalsumfor P}
\end{eqnarray}
Substituting the QCD representation of the correlation function in
Eq.(\ref{correlator of V meson}) with the help of the OPE technique,
we can derive the explicit forms of the form factors in the
light-cone QCD sum rules as
\begin{eqnarray}
V(q^2)&=&{ (m_b+m_c)(m_{B_c}+m_{V_{c \bar{c}}}) f_{V_{c \bar{c}}}
\over f_{B_c} m_{B_c}^2} {\rm{exp}} \bigg({m_{B_c}^2 \over
M^2}\bigg) \int_{\Delta}^1 {du \over u} V_{T}(u)  {\rm{exp}} \bigg[
-{m_b^2-\bar{u} (q^2-u p^2)\over u M^2}\bigg],
\nonumber \\
A_1(q^2)&=&{ (m_b+m_c) f_{V_{c \bar{c}}} \over f_{B_c} m_{B_c}^2
(m_{B_c}+m_{V_{c \bar{c}}})} {\rm{exp}} \bigg({m_{B_c}^2 \over
M^2}\bigg) \int_{\Delta}^1 {du \over u} V_{T}(u) {\rm{exp}} \bigg[
-{m_b^2-\bar{u} (q^2-u p^2)\over u M^2}\bigg] {m_b^2-q^2+u^2 p^2 \over u},
\nonumber \\
A_2(q^2)&=&-A_3(q^2)=V(q^2), \label{sum rules for V meson}
\end{eqnarray}
with the lower integral limit $\Delta$ defined by the Eq.(\ref{lower
limit}). It needs to note that similar results were also obtained in
Ref. \cite{p. ball, fen zuo}.

\subsection{Light-cone QCD sum rules for the weak transition form
factors of  $B_c \to A_{c \bar{c}}$}

The derivation of light-cone QCD sum rules for $B_c \to A_{c
\bar{c}}$ is very similar to that for $B_c \to V_{c \bar{c}}$
discussed before. The correlator for  $B_c \to A_{c \bar{c}}$ can be
given by
\begin{eqnarray}
\Pi_{\mu}(p,q)=i \int d^4 x e^{i q \cdot x} \langle A_{c
\bar{c}}(p)|T\{\bar{c}(x)\gamma_{\mu}(1-\gamma_5) b(x), \bar{b}(0) i
(1+\gamma_5) c(0)\}| 0\rangle. \label{correlator of A meson}
\end{eqnarray}
We will skip the detailed derivation of sum rules for the form
factors in $B_c \to A_{c \bar{c}}$   and only display the final
results of them as
\begin{eqnarray}
A(q^2)&=&{ (m_b+m_c)(m_{B_c}-m_{A_{c \bar{c}}}) f_{A_{c \bar{c}}}
\over f_{B_c} m_{B_c}^2} {\rm{exp}} \bigg({m_{B_c}^2 \over
M^2}\bigg) \int_{\Delta}^1 {du \over u} \phi_{\perp}(u)  {\rm{exp}}
\bigg[ -{m_b^2-\bar{u} (q^2-u p^2)\over u M^2}\bigg],
\nonumber \\
V_1(q^2)&=&{ (m_b+m_c) f_{A_{c \bar{c}}} \over f_{B_c} m_{B_c}^2
(m_{B_c}-m_{A_{c \bar{c}}})} {\rm{exp}} \bigg({m_{B_c}^2 \over
M^2}\bigg) \int_{\Delta}^1 {du \over u}\phi_{\perp}(u) {\rm{exp}}
\bigg[ -{m_b^2-\bar{u} (q^2-u p^2)\over u M^2}\bigg] {m_b^2-q^2+u^2
p^2 \over u},
\nonumber \\
V_2(q^2)&=&-V_3(q^2)=A(q^2). \label{sum rules for A meson}
\end{eqnarray}

\section{Numerical results for the form factors and decay rates}
\label{Numerical results}

Now we are going to analyze the sum rules for the form factors
numerically.  Firstly, we collect the input parameters used in this
paper as below \cite{PDG, Kiselev, korner, bauer, ioffe, Bc
lifetime}:
\begin{equation}
\begin{array}{ll}
m_e= 0.511 {\rm{MeV}}, &  m_{\tau}=1.777 {\rm{GeV}}, \\
m_b=(4.68 \pm 0.03) {\rm{GeV}}, & m_c=(1.275 \pm 0.015) {\rm{GeV}},
\\
m_{B_c}=(6.286 \pm 0.005) {\rm{GeV}}, &  f_{B_c}=(395 \pm 15)
{\rm{MeV}},
\\
\tau_{B_c}=(0.463^{+0.073}_{-0.065}) \, ps, & s_0^{B_c}=(45 \pm 1)
{\rm{GeV}}^2,
\\
G_{F}=1.166 \times 10^{-5} {\rm{GeV}}^{-2},  &  |V_{cb}|=(42.21
^{+0.10}_{-0.80}) \times 10^{-3}. \label{inputs}
\end{array}
\end{equation}
It is noted that the decay constants of various charmonium states
have been discussed comprehensively in section \ref{distribution
amplitudes}.

The choice of the threshold parameter $s$  can be determined by the
condition that the sum rules should take on the best stability in
the allowed $M^2 $ region. Besides, the value of threshold parameter
should be around the mass square of the corresponding first excited
state, hence they are also chosen the same as that in the usual
two-point QCD sum rules. The standard value of the threshold in the
$X$ channel is ${s_0}_{X}=(m_X+\Delta_X)^2$, where $\Delta_X$ is
usually taken as $0.5 \mathrm{GeV}$ \cite{dosch, matheus, bracco,
navarra} approximately in the literature. To be more specific, we
will adopt the threshold parameter for$B_c$ channel  $ s_0^{B_c}$ as
$(45 \pm 1) {\rm{GeV}}^2$ for the error estimate in the numerical
analysis as shown above.

It is well known that the form factors should not depend on the
Borel mass $M$ in the complete theory. However, we can only truncate
the operator product expansion up to some finite dimension and
perform the perturbative series in $\alpha_s$ to some order in
practice, both of which will result in the dependence of the form
factors on the Borel parameter definitely. Therefore, one should
find a region where the results only depend moderately on the Borel
mass, and the approximations for the above truncations in the
complete theory are reasonable and acceptable.

\begin{figure}[tb]
\begin{center}
\begin{tabular}{ccc}
\includegraphics[scale=0.8]{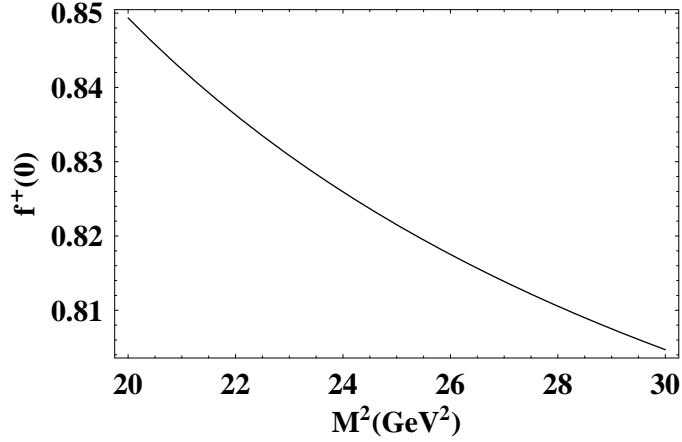}
\vspace{-2 cm}
\end{tabular}
\caption{The form factor $f_+(0)$ responsible for $B_c \to
\eta_c^{\prime}$ decay  within the Borel window.} \label{sum rules
for the P meson}
\end{center}
\end{figure}

\begin{figure}[tb]
\begin{center}
\begin{tabular}{ccc}
\includegraphics[scale=0.8]{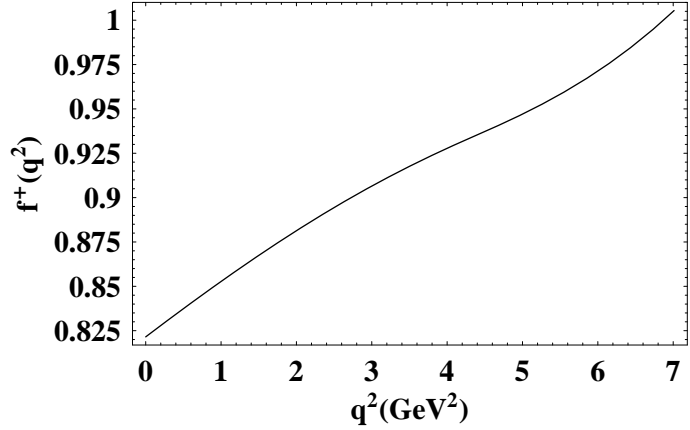}
\vspace{-2 cm}
\end{tabular}
\caption{$q^2$ dependence of the form factor $f_{+}(q^2)$ with
$M^2=25 {\rm{GeV}}^2$ in the whole physical kinematical region.}
\label{t dependence of form factor for the P meson}
\end{center}
\end{figure}

In general, the Borel mass $M$ should be chosen under the
requirement that both the contributions from the higher resonance
states and higher twist distribution amplitudes are small (no more
than 30 \%) to ensure the validity of the OPE near the light-cone
and the quark-hadron duality being a good approximation. As for the
decay of $B_c \to \eta_c^{\prime}$, we indeed find the Borel
platform $M^2 \in [20, 30] {\rm{GeV}}^2$, which is also consistent
with the number obtained in the two-point QCD sum rules
corresponding to the decay constant of $f_{B_c}$ \cite{chabab}. The
light-cone QCD sum rules of form factor $f_{+}(q^2)$ at zero
momentum transfer is shown in Fig. \ref{sum rules for the P meson}.
The values of $f_{+}(0)$ with various uncertainties rooting in Borel
mass, threshold value, decay constants of the related mesons, heavy
quark masses and the parameter $v^2$ involved in the LCDAs of
charmonium have been collected in Table~\ref{form factors of P meson
at zero momentum transfer}, from which we can find that the total
uncertainties of form factors are indeed at the level of $(20 -30)
\%$ as expected by the general understanding of the theoretical
framework. The form factor $f_{-}(0)$ up to the twist-3 LCDAs of
$P_{c \bar{c}}$ and leading order of $\alpha_s$ is zero as a result
of the large-recoil symmetry. The $q^2$ dependence of the form
factor $f_{+}(q^2)$ calculated from light cone sum rules is shown in
Fig.\ref{t dependence of form factor for the P meson} in the
physical kinematical region $0 \leq q^2 \leq (m_{B_c}
-m_{\eta_{c}^{\prime}})^2 $. Since the number of $(m_{B_c}
-m_{\eta_{c}^{\prime}})^2 \simeq 7.0 {\rm {GeV}^2} $ with
$m_{\eta_c^{\prime}}=3.638 \pm {0.004} {\rm {GeV}}$ \cite{PDG} being
used, is smaller than that of $(m_b-m_c)^2- 2 \Lambda_{QCD}
(m_b-m_c)\simeq 8.2 {\rm {GeV}^2}$, the OPE technique near the
light-cone can be performed in the whole kinematical region
effectively.

\begin{table}[tb]
\caption{The form factors $f_{i}(0)$ responsible for $B_c \to
P(S)_{c \bar{c}}$ decay in the light-cone QCD sum rules approach;
the errors for these entries correspond to the uncertainties in the
Borel mass, threshold value, quark masses, decay constants of two
mesons and variations of $v^2$ in the LCDAs of charmonium
respectively. }
\begin{center}
\begin{tabular}{|c|c|c|}
  \hline
  \hline
  decay modes & $f_{+}(0)$ & $f_{-}(0)$ \\
  \hline
  $B_c \to \eta_c^{\prime}(2^1S_0)$ & $0.82^{+0.03+0.02+0.01+0.17+0.01}_{-0.01-0.02-0.01-0.19-0.01}$ & 0 \\
  \hline
  $B_c \to X(3940)(3^1S_0)$ & $0.46^{+0.01+0.00+0.00+0.10+0.01}_{-0.01-0.01-0.01-0.11-0.01}$ & 0 \\
  \hline
  $B_c \to Y(3940)(2^3P_0)$ & $2.6^{+0.1+0.0+0.0+0.2+0.0}_{-0.1-0.1-0.1-0.2-0.2}$ & 0 \\
  \hline
\end{tabular}
\end{center}
\label{form factors of P meson at zero momentum transfer}
\end{table}

We can further evaluate the sum rules for the form factors
associating with $B_c$ to other charmonium states. For example, the
only difference for the calculation of decay  mode $B_c \to
X(3940)(3^1S_0) l \bar{\nu}_l$, is to substitute the LCDAs of $
X(3940)(3^1S_0)$ for that corresponding to $\eta_c^{\prime}$
compared with the decay of $B_c \to \eta_c^{\prime} l \bar{\nu}_l$.
In the light of Eq. (\ref{sum rules of S meson}) and the light-cone
distribution amplitudes of scalar charmonium calculated before, it's
straightforward to estimate the light-cone sum rules for the
transition form factors of $B_c \to \chi_{c0}^{\prime} l
\bar{\nu}_l$, the number of which has been grouped in
Table~\ref{form factors of P meson at zero momentum transfer}.
Evaluations of the form factors relating to the $B_c \to V(A)(c
\bar{c})$ decay are also easily carried out with the help of
Eq.(\ref{sum rules for V meson},\ref{sum rules for A meson}) and the
LCDAs of (axial) vector meson displayed in  Section
\ref{distribution amplitudes}. Since the calculations for all of
these form factors are quite similar, we will not explicitly repeat
the details anymore and only display the final results in
Table~\ref{form factors of A meson at zero momentum transfer}.


\begin{table}[tb]
\caption{The form factors $f_{i}(0)$ responsible for $B_c \to V_{c
\bar{c}} (A_{c \bar{c}})$ decays in the light-cone QCD sum rules
approach; the errors for these entries correspond to the
uncertainties in the Borel mass, threshold value, quark masses,
decay constants of two mesons and variations of $v^2$ in the LCDAs
of charmonium respectively.}
\begin{center}
\begin{tabular}{|c|c|c|}
  \hline
  \hline
  decay mode & $V(0)$ & $A_1(0)$   \\
  \hline
  $B_c \to \psi(2^3S_1)$ & $0.90^{+0.03+0.02+0.02+0.04+0.01}_{-0.02-0.03-0.02-0.05-0.00}$ & $0.38^{+0.01+0.01+0.00+0.02+0.00}_{-0.01-0.02-0.01-0.02-0.00}$\\
   \hline
  $B_c \to \psi(1^3D_1)$ & $0.11^{+0.00+0.01+0.00+0.00+0.00}_{-0.00-0.01-0.00-0.00-0.01}$ & $4.9^{+0.0+0.5+0.2+0.2+0.1}_{-0.1-0.6-0.3-0.2-0.3}\times 10^{-2}$\\
    \hline
  $B_c \to \psi(3^3S_1)$ & $0.52^{+0.02+0.00+0.00+0.04+0.00}_{-0.01-0.01-0.01-0.04-0.00}$ & $0.21^{+0.01+0.00+0.00+0.02+0.00}_{-0.01-0.00-0.00-0.02-0.00}$\\
   \hline
  $B_c \to \psi(2^3D_1)$ & $7.2^{+0.0+0.9+0.5+0.4+0.6}_{-0.1-1.0-0.6-0.5-0.6}\times 10^{-2}$ & $3.0^{+0.0+0.4+0.2+0.2+0.3}_{-0.0-0.4-0.2-0.2-0.2}\times 10^{-2}$\\
  \hline
  $B_c \to Y(4260) (4^3S_1)$ & $0.47^{+0.01+0.00+0.00+0.07+0.00}_{-0.01-0.01-0.01-0.07-0.01}$ & $0.18^{+0.01+0.01+0.00+0.03+0.01}_{-0.00-0.00-0.00-0.03-0.0}$ \\
   \hline
  $B_c \to \psi(3^3D_1)$ & $3.8^{+0.0+0.6+0.3+0.3+0.4}_{-0.1-0.7-0.4-0.3-0.5}\times 10^{-2}$ & $1.5^{+0.1+0.3+0.2+0.1+0.2}_{-0.0-0.2-0.1-0.1-0.1}\times 10^{-2}$\\
  \hline\hline
    & $A(0)$ & $V_1(0)$  \\
  \hline
  $B_c \to X(3872) (2^3P_1)$ & $-0.53^{+0.02+0.01+0.01+0.02+0.04}_{-0.03-0.00-0.00-0.02-0.03}$ & $-3.76^{+0.12+0.05+0.02+0.14+0.24}_{-0.18-0.02-0.00-0.14-0.20}$\\
  \hline
  $B_c \to X(3940) (2^3P_1)$ & $-0.51^{+0.02+0.01+0.01+0.02+0.04}_{-0.02-0.00-0.01-0.02-0.03}$ & $-3.87^{+0.11+0.05+0.02+0.15+0.24}_{-0.19-0.03-0.02-0.15-0.22}$\\
  \hline
  $B_c \to h_c(1^1P_1)$ &  $0.28^{+0.01+0.01+0.01+0.01+0.00}_{-0.01-0.00-0.00-0.01-0.00}$ & $1.51^{+0.04+0.04+0.02+0.06+0.01}_{-0.02-0.04-0.02-0.06-0.00}$\\
  \hline
  $B_c \to h_c(2^1P_1)$ &  $0.14^{+0.00+0.00+0.00+0.01+0.00}_{-0.00-0.01-0.00-0.01-0.00}$ & $1.10^{+0.00+0.00+0.04+0.00+0.00}_{-0.00-0.00-0.04-0.00-0.00}$\\
  \hline
\end{tabular}
\end{center}
\label{form factors of A meson at zero momentum transfer}
\end{table}

Utilizing the above form factors and the input parameters shown in
Eq.(\ref{inputs}), we can proceed to compute the branching ratios of
these modes. Following the standard procedure, the differential
partial decay rate for $B_c \to M_{c \bar{c}} l \bar{\nu}_{l}$ ($l=
e, \tau$) can be written as \cite{PDG}
\begin{equation}
{d\Gamma_{B_c \to M_{c \bar{c}} l \bar{\nu}_{l}} \over d q^2} ={1
\over (2 \pi)^3} {1 \over 32 m_{B_c}^3} \int_{u_{min}}^{u_{max}}
|{\widetilde{M}}_{B_c \to M_{c \bar{c}} l \bar{\nu}_{l}}|^2 du,
\end{equation}
where $u=(p_{M_{c \bar{c}}}+p_{l})^2$ and
$q^2=(p_{l}+p_{\bar{\nu}_l})^2$; $p_{M_{c \bar{c}}}$, $p_{l}$ and
$p_{\bar{\nu}_l}$ are the momenta of $M_{c \bar{c}}$, $l$ and
$\bar{\nu}_{l}$ respectively; $\widetilde{M}$ is the decay amplitude
after integrating over the angle between the $l$ and $M_{c
\bar{c}}$. The upper and lower limit of $u$ are given by
\begin{eqnarray}
u_{max}&=&(E^{\ast}_{M_{c \bar{c}}}+E^{\ast}_{l})^2-(\sqrt{E_{M_{c
\bar{c}}}^{\ast 2}-m_{M_{c \bar{c}}}^2} -\sqrt{E_l^{\ast 2}-m_l^2})^2, \nonumber\\
u_{min}&=&(E^{\ast}_{M_{c \bar{c}}}+E^{\ast}_{l})^2-(\sqrt{E_{M_{c
\bar{c}}}^{\ast 2}-m_{M_{c \bar{c}}}^{2}} +\sqrt{E_l^{\ast
2}-m_l^2})^2;
\end{eqnarray}
where $E^{\ast}_{M_{c \bar{c}}}$ and $E^{\ast}_{l}$ are the energies
of the charmonium state and the lepton in the rest frame of
lepton-neutrino pair respectively and the manifest expressions of
them can be given by
\begin{equation}
E^{\ast}_{M_{c \bar{c}}}= {m_{B_c}^2 -m_{M_{c \bar{c}}}^2 -q^2 \over
2 \sqrt{q^2}}, \qquad E^{\ast}_{l}={q^2+m_l^2 \over 2\sqrt{q^2}}.
\end{equation}
The numerical results are shown in Table~\ref{BR of Bc to electron}
and \ref{BR of Bc to charmonium tauon}, together with the numbers
obtained in other approaches  for comparison.  It is observed that
the decay rates for $B_c \to h_c$ and $B_c \to \eta_c^{\prime}$
calculated in this work are consistent with that obtained in other
frameworks \cite{korner, NRQM, ebert, c.h. chang, kiselev} within
the error bars, such as quark model, Bethe-Salpeter equation, SVZ
sum rules and so on, therefore, the branching fractions for $B_c$ to
the new charmonium states presented in this work are reliable and
acceptable.

\begin{table}[tb]
\caption{Branching fractions of $B_c \to M_{c \bar{c}} e
\bar{\nu}_e$ semi-leptonic decays in the light-cone QCD sum rules
approach; the errors for these entries correspond to the
uncertainties in the Borel mass, threshold value, quark masses,
decay constants of two mesons, lifetime of $B_c$ and variations of
$v^2$ in the LCDAs of charmonium respectively.}
\begin{center}
\begin{tabular}{|c|c|c|}
  \hline
  \hline
  decay modes & BR(this work) & Other works  \\
  \hline
  $B_c \to \eta_c^{\prime}(2^1S_0) e \bar{\nu}_e$ & $1.1^{+0.1+0.0+0.0+0.4+0.4+0.0}_{-0.0-0.0-0.0-0.5-0.4-0.0} \times 10^{-3}$ & $3.2 \times 10^{-4}$ \cite{ebert} \\
   & & $5.1 \times 10^{-4}$ \cite{c.h. chang}\\
  \hline
  $B_c \to X(3940)(3^1S_0) e \bar{\nu}_e$ & $1.9 ^{+0.2+0.1+0.0+0.8+0.7+0.0}_{-0.1-0.1-0.0-0.9-0.7-0.0} \times 10^{-4}$ &    \\
  \hline
  $B_c \to Y(3940)(2^3P_0) e \bar{\nu}_e$ & $7.2 ^{+0.9+0.1+0.1+0.9+2.5+0.0}_{-0.5-0.0-0.0-0.9-2.8-0.9} \times 10^{-3}$ &    \\
  \hline
  $B_c \to Y(4260)(4^3S_1) e \bar{\nu}_e$& $1.5 ^{+0.1+0.0+0.1+0.1+0.5+0.0}_{-0.1-0.1-0.0-0.1-0.6-0.0} \times 10^{-4}$ &    \\
  \hline
  $B_c \to X(3872)(2^3P_1) e \bar{\nu}_e$ & $6.7^{+0.9+0.0+0.1+0.5+2.3+0.7}_{-0.5-0.0-0.0-0.5-2.6-0.7} \times 10^{-3}$ &    \\
  \hline
  $B_c \to X(3940)(2^3P_1) e \bar{\nu}_e$ & $6.0^{+0.7+0.0+0.0+0.5+2.1+0.6}_{-0.5-0.1-0.1-0.5-2.3-0.7} \times 10^{-3}$ &    \\
  \hline
  $B_c \to h_c(1^1P_1) e \bar{\nu}_e$ & $2.9^{+0.3+0.0+0.0+0.2+1.0+0.1}_{-0.1-0.0-0.0-0.2-1.1-0.0} \times 10^{-3}$ & $2.7 \times 10^{-3}$ \cite{korner}
  \\ & & $1.7 ^{+0.2}_{-0.0} \times 10^{-3}$ \cite{NRQM} \\
  \hline
  $B_c \to h_c^{\prime}(2^1P_1) e \bar{\nu}_e$ & $5.3^{+0.3+0.1+0.0+0.4+1.8+0.1}_{-0.2-0.1-0.0-0.4-2.1-0.0} \times 10^{-4}$ &   \\
  \hline
\end{tabular}
\end{center}
\label{BR of Bc to electron}
\end{table}

\begin{table}[tb]
\caption{Branching fractions of $B_c \to M_{c \bar{c}} \tau
\bar{\nu}_{\tau}$ semi-leptonic decays in the light-cone QCD sum
rules approach; the errors for these entries correspond to the
uncertainties in the Borel mass, threshold value, quark masses,
decay constants of two mesons, lifetime of $B_c$ and variations of
$v^2$ in the LCDAs of charmonium respectively .}
\begin{center}
\begin{tabular}{|c|c|c|}
  \hline
  \hline
  decay modes & BR(this work) & Other works  \\
  \hline
  $B_c \to \eta_c^{\prime}(2^1S_0) \tau  \bar{\nu}_{\tau}$ & $8.1^{+0.9+0.1+0.1+3.3+2.8+0.1}_{-0.5-0.1-0.1-3.7-3.2-0.0} \times 10^{-5}$ & $1.6 \times 10^{-5}$ \cite{kiselev}   \\
  \hline
  $B_c \to X(3940)(3^1S_0) \tau \bar{\nu}_{\tau}$ & $5.7 ^{+0.6+0.7+0.3+2.4+2.0+0.0}_{-0.3-0.4-0.3-2.7-2.2-0.1} \times 10^{-6}$ &    \\
  \hline
  $B_c \to Y(3940)(2^3P_0) \tau \bar{\nu}_{\tau}$ & $2.7 ^{+0.4+0.0+0.0+0.3+0.9+0.0}_{-0.2-0.0-0.0-0.3-1.1-0.3} \times 10^{-4}$ &    \\
  \hline
  $B_c \to Y(4260)(4^3S_1) \tau \bar{\nu}_{\tau}$& $6.4^{+0.5+0.8+0.3+0.5+2.2+0.1}_{-0.3-0.4-0.2-0.5-2.5-0.0} \times 10^{-7}$ &    \\
  \hline
  $B_c \to X(3872)(2^3P_1) \tau \bar{\nu}_{\tau}$ & $3.2^{+0.5+0.0+0.0+0.2+1.1+0.4}_{-0.2-0.2-0.0-0.2-1.3-0.3} \times 10^{-4}$ &   \\
  \hline
  $B_c \to X(3940)(2^3P_1) \tau \bar{\nu}_{\tau}$ & $2.2^{+0.3+0.0+0.1+0.2+0.8+0.2}_{-0.2-0.0-0.0-0.2-0.9-0.3} \times 10^{-4}$ &    \\
  \hline
  $B_c \to h_c(1^1P_1) \tau \bar{\nu}_{\tau}$ & $3.7^{+0.4+0.1+0.1+0.3+1.3+0.1}_{-0.2-0.1-0.0-0.3-1.4-0.0} \times 10^{-4}$ &$1.7 \times 10^{-4}$ \cite{korner} \\
  & & $1.5^{+0.1}_{-0.0} \times 10^{-4}$ \cite{NRQM}     \\
  \hline
  $B_c \to h_c^{\prime}(2^1P_1) \tau \bar{\nu}_{\tau}$ & $2.0^{+0.2+0.0+0.0+0.2+0.7+0.1}_{-0.1-0.0-0.0-0.2-0.8-0.0} \times 10^{-5}$ & \\
  \hline
\end{tabular}
\end{center}
\label{BR of Bc to charmonium tauon}
\end{table}

It should be noted that the assignment of $X(3940)$ as $3^1S_0$
charmonium state leads to the production rate as $1.9 \times
10^{-4}$ in the weak decay of $B_c \to X(3940)e \bar{\nu}_{e}$, the
magnitude of which is one order smaller than that for the
interpretation of $X(3940)$ being a $2^3P_1$ charmonium. This
particular phenomenology can provide   valuable information for us
to discover the inner structures of $X(3940)$. Besides, both of
$X(3872)$ and $Y(3940)$  should be observed in the weak decay of
$B_c \to X(3872)/Y(3940)e \bar{\nu}_{e}$ in view of the branching
ratio as large as $10 ^{-3}$ order, on the condition that they can
be explained as $2^3P_1$ and $2^3P_0$ charmonium states
respectively. If future experimental measurements deviate from our
predictions heavily, it will rule out the current $J^{PC}$
assignments of the charmonium states.  It also needs to mention that
the decay rates for semi-leptonic decays of $B_c \to M_{c \bar{c}}
\tau \bar{\nu}_{\tau}$ are also displayed in Table~\ref{BR of Bc to
charmonium tauon}, from which we can find that they are about one
order smaller than the corresponding channel $B_c \to M_{c \bar{c}}
e \bar{\nu}_{e}$ due to suppression from the phrase space and
sensitive dependence of form factors on the momentum transfer $q^2$.
In particular, the branching fraction of $B_c \to Y(4260) \tau
\bar{\nu}_{\tau}$ is about two orders smaller than that for the $B_c
\to Y(4260) e \bar{\nu}_{e}$ mode, since the sum of the mass for
$Y(4260)$ and $\tau$ lepton is almost close  to the threshold of
$B_c$ meson.

Finally, we are in a position of concentrating on the $S-D$ mixing
of various vector charmonium. It's known that the $S-D$ mixing of
$\psi(3686)$ and $\psi(3770)$ may be essential to explain the large
leptonic decay width of $\psi(3770)$, the notorious $\rho \pi$
puzzle \cite{Rosner} and the enhancement of $\psi(3686) \to K_L K_S$
\cite{P. Wang}. The production of $\psi(3770)$ in $B$ meson decays
$B^{+} \to \psi(3770) K^{+}$ is found to be surprisingly large by
Belle \cite{Belle 3770}, which can be even comparable to $B^{+} \to
\psi(3686) K^{+}$  \cite {k.t. chao 1,k.t. chao 2}. Hence, it is
helpful to investigate the weak production of $\psi(3686)$ and
$\psi(3770)$ in $B_c$ decays in order to test the above mixing
scheme further and clarify the inner structures of them. Assuming
that the physical state $\psi(3686)$ and $\psi(3770)$ are the
mixture of $1^3D_1$ and $2^3S_1$ states, we have
\begin{eqnarray}
|\psi(3686)\rangle&=&{\rm{cos}} \theta | 2^3S_1\rangle + {\rm{sin}}
\theta | 1^3D_1\rangle, \nonumber \\
|\psi(3770)\rangle&=&- {\rm{sin}} \theta | 2^3S_1\rangle+{\rm{cos}}
\theta |1^3D_1\rangle.
\end{eqnarray}
As for the   mixing angle $\theta$, two solutions $\theta=-(12\pm
2)^{\rm{o}}$ or $\theta=(27 \pm 2)^{\rm{o}}$  \cite{y.p. kuang,y.b.
ding}, were found in order to reproduce the leptonic widths of
$\psi(3686)$ and $\psi(3770)$ \cite{Rosner}. The small mixing
solution, i.e., $\theta=-(12\pm 2)^{\rm{o}}$, is consistent with
couple-channel estimates \cite{Eichten,Heikkila} and the E1
transition $\psi^{\prime} \to \chi_{cJ} \gamma$ \cite{y.b. ding}.

\begin{figure}[tb]
\begin{center}
\begin{tabular}{ccc}
\includegraphics[scale=0.8]{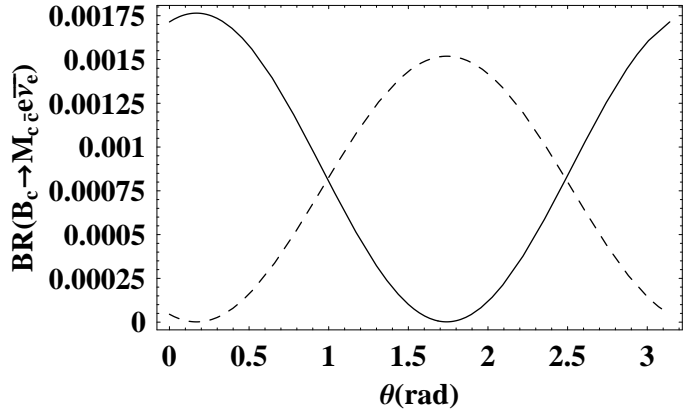}
\vspace{-2 cm}
\end{tabular}
\caption{Decay rates of $B_c \to \psi(3686) e \bar{\nu_e}$ and  $B_c
\to \psi(3770) e \bar{\nu_e}$ as functions of the mixing angle
$\theta$. The solid line represents the case of $B_c \to \psi(3686)
e \bar{\nu_e}$, while the dashed line is for $B_c \to \psi(3770) e
\bar{\nu_e}$.} \label{S-D mixing 2S}
\end{center}
\end{figure}

Based on the transition form factors of $B_c \to \psi(2^3S_1)$ and
$B_c \to \psi(1^3D_1)$ listed in Table~\ref{form factors of A meson
at zero momentum transfer}, we can plot the production rates of them
in the $B_c$ decays as functions of the mixing angle $\theta$, which
are displayed in Fig. \ref{S-D mixing 2S}. As for the favored mixing
angle $\theta=-12^{\rm{o}}$,   the branching fraction of $B_c \to
\psi(3686) e \bar{\nu_e}$ is $1.5 \times 10^{-3}$, which is almost
the same as the number of $1.7 \times 10^{-3}$   in the case  of
$2^3S_1$ state without mixing. However, the mixing component of
$\psi(2^3S_1)$ in the structure of $\psi(3770)$ is in particular
important, which can increase the decay rate of $B_c \to \psi(3770)
e \bar{\nu_e}$ from $4.5 \times 10^{-5}$ to $2.1 \times 10^{-4}$.
The reason is that the decay constant of $1^3D_1$ charmonium (47.8
\rm{MeV}) is too small compared with that of $2^3S_1$ state (304
\rm{MeV}), therefore, even a small mixing angle can affect the decay
rate of $B_c \to \psi(3770) e \bar{\nu_e}$ drastically, but almost
has no effect on the decay $B_c \to \psi(3686) e \bar{\nu_e}$.

Besides, exploring the properties of $Y(4260)$ and $\psi(4415)$ as
the mixing of $4S$ and $3D$ states in the weak decays can also shed
light on the the inner structures of these charmonium-like
particles. On the one hand, the assignment of $3^3D_1$ charmonium
for $\psi(4415)$ is supported by the fact that $\psi(4415)$ is
dominated by the decay of $\psi(4415) \to D \bar{D}_2^{\ast}(2460)$
reported by the Belle Collaboration very recently \cite{psi(4415)}.
On the other hand, $Y(4260)$ can be accommodated as $4S$ state
naturally based on the analysis of production and decay characters
of  it \cite{klempt}. Moreover, the absence of $Y(4260)$ signal in
$e^{+} e^{-} \to $ hadrons can be explained quantitatively in the
$S-D$ mixing scheme \cite{Y(4260)}. Similar to the $S-D$ mixing of
$\psi(3686)$ and $\psi(3770)$ , we express the states of $Y(4260)$
and $\psi(4415)$ as
\begin{eqnarray}
|Y(4260)\rangle&=&{\rm{cos}} \theta | 4^3S_1\rangle + {\rm{sin}}
\theta | 3^3D_1\rangle, \nonumber \\
|\psi(4415)\rangle&=&- {\rm{sin}} \theta | 4^3S_1\rangle+{\rm{cos}}
\theta |3^3D_1\rangle.
\end{eqnarray}

\begin{figure}[tb]
\begin{center}
\begin{tabular}{ccc}
\includegraphics[scale=0.8]{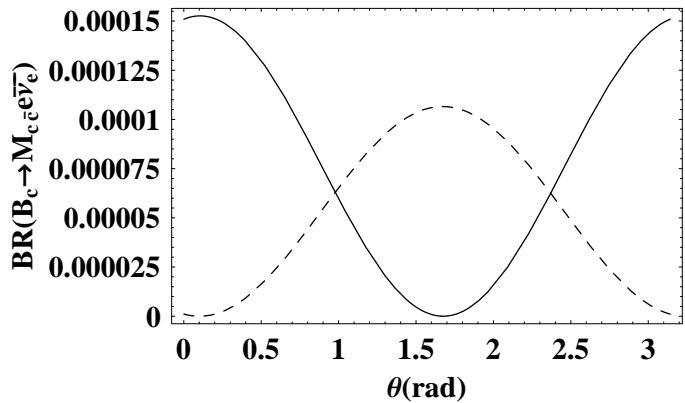}
\vspace{-2 cm}
\end{tabular}
\caption{Decay rates of $B_c \to Y(4260) e \bar{\nu_e}$ and  $B_c
\to \psi(4415) e \bar{\nu_e}$ as functions of the mixing angle
$\theta$. The solid line represents the case of $B_c \to Y(4260) e
\bar{\nu_e}$, while the dashed line is for $B_c \to \psi(4415) e
\bar{\nu_e}$.} \label{S-D mixing 4S}
\end{center}
\end{figure}

In the above mixing picture, we can analyze the dependence of
production rates for $Y(4260)$ and $\psi(4415)$ in the weak $B_c$
decays on the mixing angle $\theta$ with the help of the form
factors of $B_c \to \psi(4^3S_1)$ and $B_c \to \psi(3^3D_1)$
calculated before. As shown in Fig.~\ref{S-D mixing 4S}, it can be
observed that the decay rate for $B_c \to Y(4260) e \bar{\nu}_e$ and
$B_c \to \psi(4415) e \bar{\nu}_e$ are $1.5 \times 10^{-4}$ and $1.1
\times 10^{-6}$ for the null mixing angle. As a simple test, we find
that the production rate of $\psi(4415)$ in $B_c$ decay can reach as
large as $1.0 \times 10^{-5}$ for the mixing angle of $\theta=-12
^{\rm{o}}$; while the branching fraction of $B_c \to Y(4260) e
\bar{\nu}_e$ is  $1.4 \times 10^{-4}$, almost the same as that in
the case with zero mixing angle.

For the completeness, we also consider the $\psi(4040)$ and
$\psi(4160)$ being the mixing states of $3S$ and $2D$ as
\begin{eqnarray}
|\psi(4040)\rangle&=&{\rm{cos}} \theta | 3^3S_1\rangle + {\rm{sin}}
\theta | 2^3D_1\rangle, \nonumber \\
|\psi(4160)\rangle&=&- {\rm{sin}} \theta | 3^3S_1\rangle+{\rm{cos}}
\theta |2^3D_1\rangle.
\end{eqnarray}
The dependence of production rates for $B_c \to \psi(4040) e
\bar{\nu_e}$ and  $B_c \to \psi(4160) e \bar{\nu_e}$ on the mixing
angle $\theta$ have been plotted explicitly in Fig. {\ref{S-D mixing
3S}}.  Without the mixing of $S-D$ states, the branching ratios for
$B_c \to \psi(4040) e \bar{\nu}_e$ and $B_c \to \psi(4160) e
\bar{\nu}_e$ are $2.9 \times 10^{-4}$ and $7.1 \times 10^{-6}$
respectively. As for the mixing angle $\theta=-12 ^{\rm{o}}$, the
decay rates become $2.6 \times 10^{-4}$ and $3.3 \times 10^{-5}$ for
$B_c \to \psi(4040) e \bar{\nu}_e$ and $B_c \to \psi(4160) e
\bar{\nu}_e$, from which we can observe a large enhancement for the
production of $\psi(4160)$ as we expect.

\begin{figure}[tb]
\begin{center}
\begin{tabular}{ccc}
\includegraphics[scale=0.8]{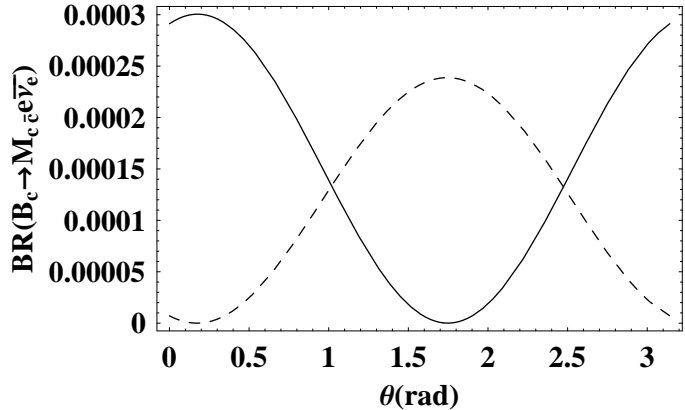}
\vspace{-2 cm}
\end{tabular}
\caption{Decay rates of $B_c \to \psi(4040) e \bar{\nu_e}$ and  $B_c
\to \psi(4160) e \bar{\nu_e}$ as functions of the mixing angle
$\theta$. The solid line represents the case of $B_c \to \psi(4040)
e \bar{\nu_e}$, while the dashed line is for $B_c \to \psi(4160) e
\bar{\nu_e}$.} \label{S-D mixing 3S}
\end{center}
\end{figure}

Up to now, we only focus on the discussions of $S-D$ mixing for the
decays of $B_c \to V_{c \bar{c}} e \bar{\nu}_e$, which can be
readily generalized to the case for the $B_c \to V_{c \bar{c}} \tau
\bar{\nu}_{\tau}$ decays. Subsequently, similar observations to the
final states being the $e \, \bar{\nu}_e$ pair can be achieved: Even
a small mixing angle can result in considerable effectrs on the
decay rates of $B_c \to \psi({n^3D_1})\tau \bar{\nu}_{\tau} \,\,
(n=1,2,3)$, while branching fractions of the corresponding channels
$B_c \to \psi({({n+1})^3S_1})\tau \bar{\nu}_{\tau}$ do not vary
significantly.

\section{Summary}

A number of new heavy charmonium  states, such as $\eta_c^{\prime}$,
$h_c$, $X(3940)$, $Y(3940)$, $X(3872)$ and $Y(4260)$ are observed
during the past several years.  There exist various explanations for
the quark components of the heavy mesons $X(3940)$, $Y(3940)$,
$X(3872)$, $Y(4260)$ so far, such as charmonium states, tetraquark
pictures, molecular bound states and so on. It is still early to
give a definite answer for their solutions.

In this work, we mainly focus on the charmonium interpretation of
all the states, $\eta_c^{\prime}$, $h_c$, $h_c^{\prime}$, $X(3940)$,
$Y(3940)$, $X(3872)$, and $Y(4260)$ produced in the exclusive
semi-leptonic weak decay of $B_c$ meson. In order to compute the
branching ratios of semi-leptonic weak decays of $B_c$, we need to
deal with the hadronic transition matrix element $\langle M_{c \bar
c}| j_{\mu} |B_c \rangle$, which defines the form factors governed
mainly by non-perturbative QCD effects. In this paper, the
light-cone QCD sum rules  approach  is used to evaluate various form
factors. We choose the correlation function with the insertion of
chiral current following the Ref. \cite{chiral current 1,chiral
current 2}, the consequences of which are that the twist-3 LCDAs do
not contribute to the sum rules for $B_c$ decays to the pseudoscalar
charmonium and also for $B_c$ decays to the (axial) vector
charmonium in the absence of three-particle wave functions.

With the help of form factors calculated in the light-cone QCD sum
rules approach, we give the decay rates for semi-leptonic decays of
$B_c \to h_c$ and $\eta_c^{\prime}$, which agree with that derived
in other frameworks. Besides, it is found that different
interpretations of $X(3940)$ can result in remarkable difference of
the production rate in the $B_c$ decays, which would help to clarify
the quark structures of the $X(3940)$ with the forthcoming LHC-b
experiments. Furthermore, the weak productions of $X(3872)$ and
$Y(3940)$ in $B_c$ decays are large enough to be detected in the
future experiments, supposing that they are indeed $2^3P_1$ and
$2^3P_0$ charmonium states respectively. It is also observed that
the mixing component of $2^3S_1$ charmonium state in the structure
of $\psi(3770)$ can enhance its production rate in $B_c$ decays
heavily, even for a small mixing angle. Besides, the production
character of $Y(4260)$ and $\psi(4415)$ as the mixing of $4S$ and
$3D$ states as well as $\psi(4040)$ and $\psi(4160)$ being the
mixing states of $3S$ and $2D$ are also included in this work. In
fact, all these decay rates depend heavily on the $J^{PC}$
assignments of the charmonium states. Therefore, our calculations
can be used in  LHC-b experiment to explore the components of these
hidden charm mesons.

\section*{Acknowledgements}

This work is partly supported by National Science Foundation of
China under Grant No.10475085 and 10625525. The authors would like
to thank C.H. Chen, V. Chernyak,  H.-n. Li,  Y. Li, Y.L.  Shen, W.
Wang, H. Zou and F. Zuo for valuable discussions.

\appendix

\section{An example of constructing  LCDAs for charmonium states}
Taking $\eta_c^{\prime \prime}$ meson as an example,  we would like
to explain the construction of LCDAs for heavy quarkonium step by
step in this appendix  based on the
procedure\cite{chernyak,hsiang-nan li} described in
section~\ref{distribution amplitudes}.

Firstly, we write down the radial Schr$\rm{\ddot{o}}$dinger
wavefunction of $n=3$, $l=0$ state for the Coulomb potential as
\begin{eqnarray}
\psi_{Sch}(r)\propto[1-{2 \over 3} q_{B}r +{2 \over 27} ( q_{B}r)^2]
{\rm exp}(-{q_B r \over 3}),
\end{eqnarray}
where $q_B$ is the Bohr momentum. Performing the Fourier
transformation of the above wavefunction, then we can arrive at
\begin{eqnarray}
\psi_{Sch}(k)\propto {q_B^4 -30 q_B^2 k^2 +81 k^4\over
(9k^2+q_B^2)^4},
\end{eqnarray}
with $k^2$ being the square of three momentum, namely $k^2 =
|{\bf{k}|^2}$. In terms of the substitution assumption
\cite{terentev}
\begin{eqnarray}
\mathbf{k}_{\perp}\rightarrow \mathbf{k}_{\perp}, \,\,\, k_{z}
\rightarrow (2x-1) {m_0 \over 2}, \,\,\, m_0^2={m_c^2 +
\mathbf{k}_{\perp}^2 \over x(1-x)}. \label{substitution assumption}
\end{eqnarray}
we should make the following replacement towards the variable $k^2$
\begin{eqnarray}
k^2 \rightarrow {\mathbf{k}_{\perp}^2 +(1-2x)^2 m_c^2 \over 4
x(1-x)}.
\end{eqnarray}
Now, we can derive the Schr$\rm{\ddot{o}}$dinger wavefunction for
$\eta_c^{\prime \prime}$ as
\begin{eqnarray}
\psi_{Sch}(x) &\propto& \int d^2 \mathbf{k}_{\perp} \psi_{Sch}(x,\mathbf{k}_{\perp}) \nonumber \\
&\propto& x(1-x)\bigg \{{x(1-x)[1-4x(1-x)(1+{v^2 \over
27})]^2\over[1-4x(1-x)(1-{v^2 \over 9})]^3}\bigg\}, \label{eta_c 3S
wavefunction}
\end{eqnarray}
where $v^2$ is defined as $v^2 =q_B^2 / m_c^2$. Following the Ref.
\cite{hsiang-nan li, chernyak}, we propose the LCDAs of
$\eta_c^{\prime \prime}$ as
\begin{eqnarray}
\phi^{v,s}(x)=\phi^{v,s}_{asy}(x) \bigg \{{x(1-x)[1-4x(1-x)(1+{v^2
\over 27})]^2\over[1-4x(1-x)(1-{v^2 \over 9})]^3}\bigg\}^{1-v^2},
\end{eqnarray}
where the power $1-v^2$ reflects the small relativistic corrections
to the Coulomb wavefunctions \cite{chernyak}. It can be observed
that above distribution amplitudes have the correct asymptotic
behavior for both the heavy quarkonium in the heavy quark limit $v^2
\to 0$ and light mesons in the $v^2 \to 1$ limit.

Moreover, it is known that the asymptotic forms of pseudoscalar
mesons can be given by
\begin{eqnarray}
\phi^{v}_{asy}(x)\propto x(1-x), \hspace {1 cm}
\phi^{s}_{asy}(x)\propto 1. \label{asymptonic forms}
\end{eqnarray}
Substituting Eq. (\ref{asymptonic forms}) into Eq. (\ref{eta_c 3S
wavefunction}), we can obtain the LCDAs for $\eta_c^{\prime \prime}$
\begin{eqnarray}
\phi^{v}(x)&=&10.8 x(1-x)\bigg \{{x (1-x) [1-4x(1-x)(1+{v^2 \over
27})]^2 \over [1-4 x(1-x)(1-{v^2 \over 9})]^3} \bigg \}^{1-v^2},
\nonumber \\
\phi^{s}(x)&=&2.1 \bigg \{{x (1-x) [1-4x(1-x)(1+{v^2 \over 27})]^2
\over [1-4 x(1-x)(1-{v^2 \over 9})]^3} \bigg \}^{1-v^2},
\end{eqnarray}
corresponding to the given value of $v^2=0.30$ as displayed in the
text, where the normalization condition $\int_0^1 \phi^{v,s}(x)
dx=1$ has been used in the derivation of above LCDAs. In addition,
we  find that the fluctuations of the phenomenological parameter
$v^2$ do not have significant effects on the shape of distribution
amplitudes for charmonium states generally within the acceptable
range $v^2 = 0.30 \pm 0.05 $, which can also be verified from the
numbers of decay rate for $B_c \to M_{c \bar{c}} l \bar{\nu}_l $
grouped in Table \ref{BR of Bc to electron} and \ref{BR of Bc to
charmonium tauon}.

\end{document}